\documentclass[aps,twocolumn,superscriptaddress,showpacs,showkeys,preprintnumbers,amsmath,amssymb]{revtex4}
\usepackage{txfonts}
\usepackage{dcolumn}
\usepackage{mathrsfs}
\usepackage{bm}
\usepackage{amsmath,amssymb,epsfig,float,graphics}

\begin{document}

\title{Radiative process of two entanglement atoms in de Sitter spacetime}
\author{Xiaobao Liu}
\affiliation{Department of
Physics, Key Laboratory of Low Dimensional Quantum Structures and
Quantum Control of Ministry of Education, and Synergetic Innovation
Center for Quantum Effects and Applications, Hunan Normal
University, Changsha, Hunan 410081, P. R. China}

\author{Zehua Tian}
\affiliation{Seoul National University, Department of Physics and Astronomy, Center for Theoretical Physics, Seoul 08826, Korea}

\author{Jieci Wang}
\affiliation{Department of
Physics, Key Laboratory of Low Dimensional Quantum Structures and
Quantum Control of Ministry of Education, and Synergetic Innovation
Center for Quantum Effects and Applications, Hunan Normal
University, Changsha, Hunan 410081, P. R. China}

\author{Jiliang Jing\footnote{Corresponding author, Email: jljing@hunn.edu.cn}}
\affiliation{Department of
Physics, Key Laboratory of Low Dimensional Quantum Structures and
Quantum Control of Ministry of Education, and Synergetic Innovation
Center for Quantum Effects and Applications, Hunan Normal
University, Changsha, Hunan 410081, P. R. China}

\begin{abstract}
We investigate the radiative processes of a quantum system composed by two identical two-level atoms in the de Sitter spacetime, interacting with a conformally coupled massless scalar field prepared in the de Sitter-invariant vacuum. We discuss the structure of the rate of variations of the atomic energy for two static atoms. Following a procedure developed by Dalibard, Dupont-Roc and Cohen-Tannoudji, our intention is to identify in a quantitative way the contributions of vacuum fluctuations and the radiation reaction to the generation of quantum entanglement and to the degradation of entangled states.
We find that when the distance between two atoms larger than the characteristic length scale, the rate of variation of atomic energy in the de Sitter-invariant vacuum behaves differently compared with that in the thermal Minkowski spacetime. In particular, the generation and degradation of quantum entanglement can be enhanced or inhibited, which are dependent not only on the specific entangled state but also on the distance between the atoms.
\end{abstract}

\maketitle

\section{introduction}
Superposition of quantum states and quantum entanglement are the most peculiar features of
quantum systems arguably, without analogs in classical mechanics \cite{Horodecki}.
When global states of a composite system cannot be factorized into a product of the states of their individual subsystems, it is known as quantum entanglement, exhibiting the nonlocal nature of quantum mechanics. Moreover, quantum entanglement is an important physical resource, which is at
the basis of many quantum information processing tasks such as
quantum teleportation \cite{Bennett0} and dense coding \cite{Bennett1}.
There exist several works to generate entangled two-level systems interacting with a bosonic field as shown in Refs. \cite{Plenio,Bachanov,Ficek0,Tan,Amico,Tian2013,Tian2014}.

On the other hand, spontaneous emission and excitation are one of the most important features of atoms, which can be attributed to vacuum fluctuations \cite{Welton,Compagno}, or the radiation reaction \cite{Ackerhalt}, or a combination of them \cite{Milonni1,Milonni2}. In this respect, radiative processes of entangled states have been sufficiently investigated in Refs. \cite{Guo,Ficek,Yu,Agarwal}. Recently, reference \cite{Arias} presented a heuristic scenario to consider radiative processes of entangled two-level atoms which coupled with a massless scalar field prepared in the vacuum state in the presence of boundaries, investigating the spontaneous transition rates from the entangled states to their collective ground state which is triggered by vacuum fluctuations of the quantum field. Nevertheless, following the seminal work of Ackerhalt \emph{et al.} \cite{Ackerhalt}, it is possible to interpret spontaneous decay as a radiation-reaction effect. As carefully demonstrated by Milonni in Ref.~\cite{Milonni0}, both effects of vacuum fluctuations and the radiation reaction are dependent on a particular ordering chosen for commuting and field operators. Subsequently, Dalibard \emph{et al.} developed the Dalibard-Dupont-Roc-Cohen-Tannoudji (DDC) formalism, asserting that one can distinctively separate the contributions of vacuum fluctuations and the radiation reaction as having an independent physical meaning, if a symmetric ordering of the commuting atom and field variables is adopted~\cite{Dalibard1,Dalibard2}. This is a consequence of the fact that within such a formalism the atom is stable in its ground state due to the balance between vacuum fluctuations and the radiation reaction.
We remark that in recent investigations regarding quantum entanglement, the DDC formalism was employed to understand the radiative processes involving entangled atoms in Minkowski  and Schwarzschild spacetime \cite{Menezes1,Menezes2,Menezes3}. Moreover, the DDC formalism has been proved to substantially implemented in many physical situations \cite{Audretsch0,Audretsch1,Zhu0,Rizzuto,Yu0,Yu1,Yu2,Yu3}.
There also exist related works on the resonance interaction between entangled atoms in the literatures \cite{Rizzuto1,Zhou1,Zhou2,Zhou3,Tian,Tian1}.

De Sitter space is the unique maximally symmetric curved spacetime which enjoys the degree of symmetry similar to the Minkowski spacetime and has ten Killing vectors. More importantly, our current observations and the theory of inflation suggest that our universe may approach the de Sitter geometries in the far past and the far future. And a holographic duality may exist between quantum
gravity on the de Sitter space and a conformal field theory living on the boundary identified with the
timelike infinity of de Sitter space \cite{Strominger}. In the present paper, we focus our attention on applying the DDC formalism to investigate the quantum entanglement of atoms in de Sitter spacetime. Being more specific, we consider that both atoms interact with a conformally coupled massless field in the de Sitter invariant vacuum. In addition, we intend to study the contributions of vacuum fluctuations and the radiation reaction in the radiation process of the entangled states.

The organization of the paper is as follows. In sec. II we discuss the implementation of the DDC formalism in the situation of interest. In sec. III, we calculate in detail the rate of variation of the atomic energy in de Sitter invariant vacuum when both atoms are at rest, and compare it with the scenario involving the Minkowski spacetime with a field in a thermal state. Finally, a summary of the main results of our work is present in Sec. IV. In this paper, we use units such that  $c = \hbar = G = k_B = 1$.

\section{two identical atoms coupled with a conformally coupled massless scalar field in de Sitter spacetime}
In this section we assume that two identical two-level atoms labeled by 1 and 2, respectively, are in interaction with a conformally coupled massless field in de Sitter spacetime.
It is well know that the de Sitter metric is a solution of the Einstein equations with the cosmological constant $\Lambda$. Four-dimensional de Sitter spacetime is most easily represented as the hyperboloid
\begin{eqnarray}\label{metric0}
z_0^2-z_1^2-z_2^2-z_3^2-z_4^2=-\alpha^2,
\end{eqnarray}
which is embedded in the five dimensional Minkowski spacetime with the metric
\begin{eqnarray}\label{metric}
ds^2=dz_0^2-dz_1^2-dz_2^2-dz_3^2-dz_4^2,
\end{eqnarray}
where $\alpha=\sqrt{3/\Lambda}$.  There are several different coordinate systems that can be
chosen to characterize the de Sitter spacetime~\cite{Gibbons,Birrell,Mottola}.
If we choose \emph{the static coordinates system} $(t,r,\theta,\phi)$, which is
\begin{eqnarray}
z_0&=&\sqrt{\alpha^2-r^2}\sinh{t/\alpha},\nonumber\\
z_1&=&\sqrt{\alpha^2-r^2}\cosh{t/\alpha},\nonumber\\
z_2&=&r\cos{\theta},\nonumber\\
z_3&=&r\sin{\theta}\cos{\phi},\nonumber\\
z_4&=&r\sin{\theta}\sin{\phi},
\end{eqnarray}
then the corresponding line element becomes
\begin{eqnarray}
ds^2=\bigg(1-\frac{r^2}{\alpha^2}\bigg)dt^2-\bigg(1-\frac{r^2}{\alpha^2}\bigg)^{-1}
dr^2-r^2(d\theta^2+\sin^2\theta d\phi^2).\nonumber\\
\end{eqnarray}
Noting that the cosmological horizon is located at $r=\alpha$, the position in the de Sitter coordinate becomes singular.

Now let us identify the distinct contributions of vacuum fluctuations and the radiative reaction to entanglement dynamics of atoms in de Sitter spacetime. With this regard, we will work within the Heisenberg picture. Let us consider both atoms following different stationary trajectories $x^\mu(\tau_a)$, where $\tau_a$ denotes the proper time of the atom $a$ with $a=1,2$. Hereafter we describe the time evolution of the total system is to be taken with respect to the proper time $\tau$ of the atoms.
It is worth to note that the stationary trajectory guarantees the existence of stationary states. Within the multipolar coupling approach, the purely atomic part of the total Hamiltonian describes the free Hamiltonian of two identical atoms. Then the Hamiltonian of the two atoms can be written as
\begin{eqnarray}\label{H_A}
H_A(\tau)=\frac{\omega_0}{2}\bigg[(\sigma^{1}_3(\tau)\otimes \hat{1})+(\hat{1}\otimes \sigma^{2}_3(\tau))\bigg],\nonumber\\
\end{eqnarray}
where $\sigma^{a}_3=|e_a\rangle\langle e_a|-|g_a\rangle\langle g_a|$, $a=1,2$. Here, $|g_1\rangle$ and $|g_2\rangle$ are the ground states of the isolated atoms with energies $-\omega_0/2$, and $|e_1\rangle$ and $|e_2\rangle$ are the respective exited states with energies $+\omega_0/2$. One has that the space of the two-atom system is spanned by four product stationary states which are eigenstates of $H_A$ with respective energies \cite{Ficek1,Ficek2}
\begin{eqnarray}\label{product states}
E_{gg}&=&-\omega_0,\;\;\;|gg\rangle=|g_1g_2\rangle,\nonumber\\
E_{ge}&=&0,\;\;\;\;\;\;\;|ge\rangle=|g_1e_2\rangle,\nonumber\\
E_{eg}&=&0,\;\;\;\;\;\;\;|eg\rangle=|e_1g_2\rangle,\nonumber\\
E_{ee}&=&\omega_0,\;\;\;\;\;|gg\rangle=|e_1e_2\rangle.
\end{eqnarray}
In this paper, we can conveniently take the Bell state basis instead of the above product-state basis, which is
\begin{eqnarray}\label{Bell states}
&&|\psi^{\pm}\rangle=\frac{1}{\sqrt{2}}(|g_1e_2\rangle \pm |e_1g_2\rangle),\nonumber\\
&&|\phi^{\pm}\rangle=\frac{1}{\sqrt{2}}(|g_1g_2\rangle \pm |e_1e_2\rangle).
\end{eqnarray}
It is known that such states constitute familiar examples of four maximally entangled two-qubit Bell states, and form a convenient basis of the two-qubit Hilbert space.
However, we note that the Bell states $|\phi^{\pm}\rangle$ are not eigenstates of the atomic Hamiltonian $H_A$.
Here we consider the interaction between the two-atom system and a conformally coupled massless field.
We can obtain the free Hamiltonian of the quantum scalar field which is
\begin{eqnarray}
H_F(\tau)=\int d k \omega_k a^\dagger_k(t(\tau))a_k(t(\tau))\frac{dt}{d\tau},
\end{eqnarray}
where $a^\dagger_{k}$ and $a_{k}$ denote the creation and annihilation operators with momentum $k$ and $dt/d\tau=|g_{00}|^{-1/2}=(1-r^2/\alpha^2)^{-1/2}$. In addition, we have neglected the zero-point energy.
Hence one can obtain the Hamiltonian which describes the interaction between the atoms and the field
\begin{eqnarray}
H_I(\tau)=\lambda m_{1}(\tau)\varphi(x_1(\tau))
+ \lambda m_{2}(\tau)\varphi(x_2(\tau)),
\end{eqnarray}
where $\lambda$ is the coupling constant that we assume to be small and $\varphi(x)$ is the scalar field operator in de Sitter spacetime. $m_a(\tau)$ is the monopole moment operator for the $a$-th atom
\begin{eqnarray}
m_a(\tau)=\sigma_a^+(\tau)+\sigma_a^-(\tau),
\end{eqnarray}
where $\sigma_a^+=|e_a\rangle\langle g_a|$ and $\sigma_a^-=|g_a\rangle\langle e_a|$ are the atomic raising and lowering operators, respectively.

We first solve the Heisenberg equations of motion for the dynamical variable of the atoms and the field with respect to $\tau$ which can be derived from the total Hamiltonian $H(\tau)=H_A(\tau)+H_{F}(\tau)+H_I(\tau)$. Therefore, the solutions of the equation of motion can be separated in two parts, namely, the \emph{free part} which is the absence of the coupling between the atoms and fields, and the \emph{source part} which is caused by the interaction between atoms and fields. Therefore, one can obtain the atomic operators
\begin{eqnarray}
&&\sigma_3^{a}(\tau)=\sigma_3^{a,f}(\tau)+\sigma_3^{a,s}(\tau),\nonumber\\
&&m_{a}(\tau)=m^{f}_a(\tau)+m^{s}_a(\tau),
\end{eqnarray}
and the field operators
\begin{eqnarray}\label{field operators}
a_k(t(\tau))=a_k^f(t(\tau))+a_k^s(t(\tau)),
\end{eqnarray}
where the superscripts $``f"$ and $``s"$ stand for the free and source parts, respectively.
Therefore, one can also construct
$\varphi(x_a(\tau))=\varphi^f(x_a(\tau))+\varphi^s(x_a(\tau))$.
Nonetheless, there arises an ambiguity of operator ordering problem which indicates that one must choose an operator ordering when discussing the effects of vacuum fluctuations (which are originated from $\varphi^f$) and the radiation reaction (which is caused by $\varphi^s$). In this way,
it allows us to interpret the effects of such phenomena as independent physical meanings \cite{Dalibard1,Dalibard2,Audretsch0}.
Following the above prescription, let us present the contributions of quantum vacuum fluctuations and the radiation reaction in the evolution of the atoms' energies, which are given by the expectation value of $H_A$, given by Eq.~(\ref{H_A}). By employing a perturbative treatment, we take into account only terms up to order $\lambda^2$. Furthermore, we perform an averaging over the field degrees of freedom by taking vacuum expectation values. In turn, since we are interested in the evolution of expectation values of atomic observables, we also take the expectation value in the atoms' state $|\omega\rangle$, with energy $\omega$. Such a state can be one of the product states given by Eq.~(\ref{product states}) or the Bell states Eq.~(\ref{Bell states}).
Therefore, one has the vacuum-fluctuation and radiation-reaction contributions to the rate of change of $\langle H_A\rangle$, respectively,
\begin{eqnarray}\label{both contributions3}
\bigg\langle \frac{d H_A}{d\tau} \bigg\rangle_{VF}=\frac{i \lambda^2}{2}\int^{\tau}_{\tau_0}d\tau'\sum^2_{a,b=1}
D_{ab}(x_a(\tau),x_b(\tau'))\frac{d}{d\tau}\Delta^{ab}(\tau,\tau')\;,\nonumber\\
\bigg\langle \frac{d H_A}{d\tau} \bigg\rangle_{RR}=\frac{i \lambda^2}{2}\int^{\tau}_{\tau_0}d\tau'\sum^2_{a,b=1}
\Delta_{ab}(x_a(\tau),x_b(\tau'))\frac{d}{d\tau}D^{ab}(\tau,\tau')\;,\nonumber\\
\end{eqnarray}
where $a,b=1,2$, and $\langle\cdot\cdot\cdot\rangle=\langle 0,\nu|\cdot\cdot\cdot|0,\nu\rangle$ with $|0\rangle$ being the field in the de Sitter invariant vacuum state, and the subscripts ``VF" and ``RR" stand for the vacuum-fluctuation and radiation-reaction contribution, respectively.
In the above equations, the statistical functions of the field, $D_{ab}$ and $\Delta_{ab}$, are defined as
\begin{eqnarray}\label{field1}
D_{ab}(x_a(\tau),x_b(\tau'))&=&\langle 0|\{\varphi^{f}(x_a(\tau)),\varphi^{f}(x_b(\tau'))\}|0\rangle,
\end{eqnarray}
and
\begin{eqnarray}\label{field2}
\Delta_{ab}(x_a(\tau),x_b(\tau'))&=&\langle 0|[\varphi^{f}(x_a(\tau)),\varphi^{f}(x_b(\tau'))]|0\rangle,
\end{eqnarray}
where $D_{ab}$ is called the Hadamard elementary function of the field, $\Delta_{ab}$ is the Pauli-Jordan function of the field, as well as $\{,\}$ and $[,]$ denote the anti-commutator and the commutator, respectively. Besides, they are also expressed in term of the statistical functions of the free part of the atom's variable,
\begin{eqnarray}\label{atom1}
\Delta^{ab}(\tau,\tau')=\langle\omega|[m^{f}_a(\tau),m^{f}_b(\tau')]|\omega\rangle,
\end{eqnarray}
and
\begin{eqnarray}\label{atom2}
D^{ab}(\tau,\tau')=\langle \omega|\{m^{f}_a(\tau),m^{f}_b(\tau')\}|\omega\rangle,
\end{eqnarray}
where  $\Delta^{ab}$ and $D^{ab}$ are the linear susceptibility and the symmetric correlation function of the two-atom system in the state $|\omega\rangle$, respectively.

It is worth noting that $\Delta^{ab}$ and $D^{ab}$ are only characterized by the two-atom system itself, which is not dependent on the trajectory of the atoms.
In Eq.~(\ref{both contributions3}), it implies that the two contributions in the rate of the atoms' energies not only are related with  the isolated atoms, and also are due to the cross correlations
between the atoms mediated by the field. This interference attributes to the interaction of each atom with field.
Such observations show that such a formalism can make it possible for us to investigate the interaction between the vacuum fluctuations and the
radiation reaction in the entanglement generation or degradation between atoms.

According to the completeness relation $\sum_{\omega'}|\omega'\rangle\langle\omega'|=1$ and
$m_a^f(\tau)=e^{i H_A \tau}m_a^f(0)e^{-i H_A \tau}$,
one can obtain explicit forms of statistical functions for the two-atom system which are given by
\begin{eqnarray}\label{linear susceptibility}
\Delta^{ab}(\tau,\tau')=\sum_{\omega'}\bigg[\mathcal{M}_{ab}(\omega,\omega')e^{i \triangle\omega(\tau-\tau')}-\mathcal{M}_{ba}(\omega,\omega')e^{-i \triangle\omega(\tau-\tau')}\bigg],\nonumber\\
\end{eqnarray}
and
\begin{eqnarray}\label{symmetric correlation}
D^{ab}(\tau,\tau')=\sum_{\omega'}\bigg[\mathcal{M}_{ab}(\omega,\omega')e^{i \triangle\omega(\tau-\tau')}+\mathcal{M}_{ba}(\omega,\omega')e^{-i \triangle\omega(\tau-\tau')}\bigg],\nonumber\\
\end{eqnarray}
where  $\triangle\omega=\omega-\omega'$ with $\omega'$ being the energy of the state $|\omega'\rangle$, and the sum extends over the complete set of atomic states. In addition, the atomic transition monopole moment $\mathcal{M}_{ab}(\omega,\omega')$ is defined by
\begin{eqnarray}\label{monopole moment}
\mathcal{M}_{ab}(\omega,\omega')=\langle\omega|m^{f}_a(0)|\omega'\rangle\langle\omega'|m^{f}_b(0)|\omega\rangle\;.
\end{eqnarray}

In this paper we are interested in transitions from entangled states [currently represented by the Bell states] to one of the separable states, or the inverse, where the separable states and the Bell states are given by Eqs.~(\ref{product states}) and (\ref{Bell states}), respectively. On the one hand, to study the entanglement degradation between atoms as a spontaneous emission phenomenon, we suppose the atoms prepared in an entangled state $|\psi^{\pm}\rangle$. Therefore, according to Eqs.~(\ref{product states}) and (\ref{Bell states}), the only allowed transitions are $|\psi^{\pm}\rangle\rightarrow|gg\rangle$, with the energy gap $\Delta\omega=\omega-\omega'=\omega_0>0$, as well as $|\psi^{\pm}\rangle\rightarrow|ee\rangle$,
with $\Delta\omega=\omega-\omega'=-\omega_0<0$. On the other hand, if one wishes to investigate the entanglement generation between atoms, we assume that the atoms were initially prepared in the atomic separate state. When the atoms prepared in the ground state $|gg\rangle$ initially, only the transition $|gg\rangle\rightarrow|\psi^{\pm}\rangle$, with $\Delta\omega=\omega-\omega'=-\omega_0<0$, is permitted. Similarly, when the atoms are initially prepared in the excited state, we have $|ee\rangle\rightarrow|\psi^{\pm}\rangle$ with $\Delta\omega=\omega-\omega'=\omega_0>0$. We also note that if one considers the transition from the $|gg\rangle$ or $|ee\rangle$ to  $|\phi^{\pm}\rangle$, we get $\Delta\omega=\omega-\omega'=0$. This implies that the rate of variation of atomic energy is zero, i.e., $\langle d H_A/d\tau\rangle=0$. Hence it is not necessary to consider the generation of such Bell states out of the separate states.

\section{rate of variation of the atomic energy for static atoms}
In this section let us consider our static two-atom system in a situation where both atoms move along different trajectories
$r_1=r, \theta_1=\theta, \phi_1=\phi$ and $r_2=r, \theta_2=\theta', \phi_2=\phi$, respectively. We are interested in the de Sitter-invariant vacuum state which describes the conformally coupled massless scalar field, due to the fact that it is an analogous state to the Minkowski vacuum in flat spacetime \cite{Allen}. Then we will calculate the rate of variation of atomic energy for the atoms in de Sitter spacetime interacting with a conformally coupled massless scalar field in the de Sitter-invariant vacuum. We discuss the permissible transition as considered at the end of the previous section. The relevant correlation function of the conformally coupled massless scalar field which appears in Eq.~(\ref{both contributions3}) is given by
\begin{eqnarray}\label{wight function}
G^+(x,x')=-\frac{1}{4\pi^2}\frac{1}{(z_0-z_0')^2-\triangle z^2-i\epsilon},
\end{eqnarray}
where $\triangle z^2=(z_1-z_1')^2+(z_2-z_2')^2+(z_3-z_3')^2+(z_4-z_4')^2$ and $\epsilon$ is an infinitesimal constant.
According to the trajectory of two-atom system and Eq.~(\ref{wight function}), the associated Hadamard's elementary functions of the field, $D_{ab}$ , can be calculated\begin{widetext}
\begin{eqnarray}\label{D}
&&D_{ab}(x_a(\tau),x_b(\tau'))
=-\frac{1}{16\pi^2}\bigg[\frac{1}{\kappa^2 \sinh^2(\frac{\tau-\tau'}{2\kappa}-i\epsilon)}+\frac{1}{\kappa^2 \sinh^2(\frac{\tau'-\tau}{2\kappa}-i\epsilon)}\bigg],\;\;\;\;\;\;\;\;\;\;\;
\;\;\;\;\;\;\;\;\;\;\;\;\;\;\;\;\;\;\;\;\;\;\;\;\;\;\;\;\;\;\;\;\;(a=b),\nonumber\\
&&D_{ab}(x_a(\tau),x_b(\tau'))
=-\frac{1}{16\pi^2}\bigg[\frac{1}{\kappa^2 \sinh^2(\frac{\tau-\tau'}{2\kappa}-i\epsilon)-r^2\sin^2(\frac{\theta-\theta'}{2})}+\frac{1}{\kappa^2 \sinh^2(\frac{\tau'-\tau}{2\kappa}-i\epsilon)-r^2\sin^2(\frac{\theta'-\theta}{2})}\bigg],\;\;(a\neq b),
\end{eqnarray}
where $\kappa=\sqrt{g_{00}}\alpha=\sqrt{1-r^2/\alpha^2}\alpha$ and $\tau-\tau'=\sqrt{g_{00}}(t-t')$.
Applying the statistical functions of the field Eq.~(\ref{D}) and the statistical functions of the atom Eq.~(\ref{linear susceptibility}) into the Eq.~(\ref{both contributions3}), the contribution of the vacuum-fluctuation
to the rate of variation of the atomic energy can be evaluated.
Performing a change of variable $\mu=\tau-\tau'$, with $\Delta \tau=\tau-\tau_0$, we have
\begin{eqnarray}\label{vacuum fluctuations0}
\bigg\langle \frac{d H_A}{d\tau} \bigg\rangle_{VF}&=&\frac{\lambda^2}{32\pi}\sum_{\omega'}\triangle\omega\int^{\bigtriangleup \tau}_{-\bigtriangleup \tau}du \bigg\{\bigg(\mathcal{M}_{11}(\omega,\omega')+\mathcal{M}_{22}(\omega,\omega')\bigg)
\bigg[\frac{1}{\kappa^2 \sinh^2(\frac{u}{2\kappa}-i\epsilon)}+\frac{1}{\kappa^2 \sinh^2(\frac{u}{2\kappa}+i\epsilon)}\bigg]e^{i \triangle\omega u}\nonumber\\
&+&\bigg(\mathcal{M}_{12}(\omega,\omega')+\mathcal{M}_{21}(\omega,\omega')\bigg)
\bigg[\frac{1}{\kappa^2 \sinh^2(\frac{u}{2\kappa}-i\epsilon)-r^2\sin^2(\frac{\Delta\theta}{2})}
+\frac{1}{\kappa^2 \sinh^2(\frac{u}{2\kappa}+i\epsilon)-r^2\sin^2(\frac{\Delta\theta}{2})}\bigg]e^{i \triangle\omega u}\bigg\},
\end{eqnarray}
\end{widetext}
where  $\epsilon\rightarrow0^{+}$ and $\Delta\theta=\theta-\theta'$. Through the contour integral, we may compute the relevant integrals. For sufficiently long times $\bigtriangleup \tau\rightarrow\infty$, one has
\begin{eqnarray}\label{vacuum fluctuations1}
\bigg\langle \frac{d H_A}{d\tau} \bigg\rangle_{VF}=&-&\frac{\lambda^2}{4\pi}\sum_{a,b=1}^2\bigg[\sum_{\omega>\omega'}
\mathcal{M}_{ab}(\omega,\omega')(\triangle\omega)^2\nonumber\\
&\times&\bigg(1+\frac{2}{e^{2\pi\kappa\triangle \omega}-1}\bigg)
f_{ab}(\triangle \omega,L/2)\nonumber\\
&-&\sum_{\omega<\omega'}\mathcal{M}_{ab}(\omega,\omega')(\triangle\omega)^2\nonumber\\
&\times&\bigg(1+\frac{2}{e^{2\pi\kappa|\triangle \omega|}-1}\bigg)
f_{ab}(\triangle \omega,L/2)\bigg],
\end{eqnarray}
where we have defined
\begin{eqnarray}
f_{ab}(\triangle \omega,L/2)=
\begin{cases}
\;1,\;\;\;\;\;\;\;\;\;\;\;\;\;\;\;\;\;\;\;\;\;\;\;\;\;(a=b)\;,\\
\\
\;\frac{\sin[2\kappa\triangle \omega\sinh^{-1}(L/2\kappa)]}{L\triangle \omega\sqrt{1+L^2/(2\kappa)^2}},\;\;(a\neq b)\;,
\end{cases}
\end{eqnarray}
\emph{with $ L=2r\sin(\Delta\theta/2)$ being the distance between the two atoms in the 5-dimensional embedding space (\ref{metric0})-(\ref{metric}).}
It becomes clear to note that when only the first term ($\omega>\omega'$) contributes, i.e., $\triangle\omega>0$, we have $\langle d H_A/d\tau\rangle_{VF}<0$, which means that the vacuum fluctuations lead to a deexcitation of the two-atom system.
However, when $\triangle\omega<0$ ($\omega<\omega'$), there is only contribution from the second term. One has $\langle d H_A/d\tau\rangle_{VF}>0$ which implies the excitation of two-atom system. This is reminiscent from the fact that the stimulated excitation and deexcitation have equal Einstein B coefficients when an atom interacting with quantized radiation, which indicates that vacuum fluctuations will excite an atom in the ground state as well as deexcite an atom in the excited state \cite{Milonni}. As discussed at the end of the previous section, this tells us that vacuum fluctuations can  promote the entanglement degradation or the entanglement generation between the atoms.

On the other hand, applying the trajectory of two-atom system and Eq.~(\ref{wight function}) into the equation $\Delta_{ab}$, the Pauli-Jordan functions of the field are given by
\begin{widetext}
\begin{eqnarray}
&&\Delta_{ab}(x_a(\tau),x_b(\tau'))
=-\frac{1}{16\pi^2}\bigg[\frac{1}{\kappa^2 \sinh^2(\frac{\tau-\tau'}{2\kappa}-i\epsilon)}-\frac{1}{\kappa^2 \sinh^2(\frac{\tau'-\tau}{2\kappa}-i\epsilon)}\bigg],\;\;\;\;\;\;\;\;\;\;\;\;\;\;\;\;\;\;(a=b),\nonumber\\
&&\Delta_{ab}(x_a(\tau),x_b(\tau'))\nonumber\\
&&=-\frac{1}{16\pi^2}\bigg[\frac{1}{\kappa^2 \sinh^2(\frac{\tau-\tau'}{2\kappa}-i\epsilon)-r^2\sin^2(\frac{\theta-\theta'}{2})}
-\frac{1}{\kappa^2 \sinh^2(\frac{\tau'-\tau}{2\kappa}-i\epsilon)-r^2\sin^2(\frac{\theta'-\theta}{2})}\bigg],\;\;(a\neq b).
\end{eqnarray}
Performing similar calculations as above, one has the contribution of the radiation reaction
\begin{eqnarray}\label{radiation reaction0}
\bigg\langle \frac{d H_A}{d\tau} \bigg\rangle_{RR}&=&\frac{\lambda^2}{32\pi}\sum_{\omega'}\triangle\omega\int^{\bigtriangleup \tau}_{-\bigtriangleup \tau}du \bigg\{\bigg(\mathcal{M}_{11}(\omega,\omega')+\mathcal{M}_{22}(\omega,\omega')\bigg)
\bigg[\frac{1}{\kappa^2 \sinh^2(\frac{u}{2\kappa}-i\epsilon)}-\frac{1}{\kappa^2 \sinh^2(\frac{u}{2\kappa}+i\epsilon)}\bigg]e^{i \triangle\omega u}\nonumber\\
&+&\bigg(\mathcal{M}_{12}(\omega,\omega')+\mathcal{M}_{21}(\omega,\omega')\bigg)
\bigg[\frac{1}{\kappa^2 \sinh^2(\frac{u}{2\kappa}-i\epsilon)-r^2\sin^2(\frac{\Delta\theta}{2})}
-\frac{1}{\kappa^2 \sinh^2(\frac{u}{2\kappa}+i\epsilon)-r^2\sin^2(\frac{\Delta\theta}{2})}\bigg]e^{i \triangle\omega u}\bigg\}.
\end{eqnarray}
\end{widetext}
By invoking the method of residue, we extend the range of integration to infinity for $\triangle \tau\rightarrow \infty$ and the result is
\begin{eqnarray}\label{radiation reaction1}
\bigg\langle \frac{d H_A}{d\tau} \bigg\rangle_{RR}=&-&\frac{\lambda^2}{4\pi}\sum_{a,b=1}^2\bigg[\sum_{\omega>\omega'}
\mathcal{M}_{ab}(\omega,\omega')(\triangle\omega)^2f_{ab}(\triangle \omega,L/2)\nonumber\\
&+&\sum_{\omega<\omega'}\mathcal{M}_{ab}(\omega,\omega')(\triangle\omega)^2f_{ab}(\triangle \omega,L/2)\bigg].
\end{eqnarray}
We note that the effect of the radiation reaction is to induce a loss of atomic energy $\langle d H_A/d\tau\rangle_{RR}<0$ which is independent of how the qubit was initially prepared. In other words, the radiation reaction does not contribute to the entanglement generation between the atoms through an absorption process. Besides, it leads always to disentanglement via spontaneous emission processes.
This is reminiscent from the fact that for an entangled two-level system interacting with classical noise. Such noise will generally tend to decoherence and disentanglement processes.

Adding up the contributions of vacuum fluctuation and the reaction given by Eqs.~(\ref{vacuum fluctuations1}) and (\ref{radiation reaction1}), the total rate of change of the atomic energy can be obtained
\begin{eqnarray}\label{total rate0}
\bigg\langle \frac{d H_A}{d\tau} \bigg\rangle_{tot}=
&-&\frac{\lambda^2}{2\pi}\sum_{a,b=1}^2\bigg[\sum_{\omega>\omega'}\mathcal{M}_{ab}(\omega,\omega')
(\triangle\omega)^2\nonumber\\
&\times&\bigg(1+\frac{1}{e^{2\pi\kappa\triangle \omega}-1}\bigg)f_{ab}(\triangle \omega,L/2)\bigg]\nonumber\\
&+&\frac{\lambda^2}{2\pi}\sum_{a,b=1}^2\bigg[\sum_{\omega<\omega'}\mathcal{M}_{ab}(\omega,\omega')
(\triangle\omega)^2\nonumber\\
&\times&\frac{1}{e^{2\pi\kappa|\triangle \omega|}-1}f_{ab}(\triangle \omega,L/2)\bigg].\nonumber\\
\end{eqnarray}
It clearly shows that the delicate balance between the contributions of
vacuum fluctuations and the radiation reaction is broken. Therefore, the transitions to
higher levels states from ground state are possible. In the present context, it is possible to generate entanglement between the atoms via absorption processes for large asymptotic times.
In addition, from Eqs.~(\ref{product states}) and (\ref{Bell states}),
as discussed above, the nonzero monopole matrix elements in Eq.~(\ref{monopole moment}) are given by
\begin{eqnarray}
&&\mathcal{M}_{11}(\omega,\omega')=\frac{1}{2}\;,\;\;\nonumber\\
&&\mathcal{M}_{22}(\omega,\omega')=\frac{1}{2}\;,\nonumber\\
&&\mathcal{M}_{12}(\omega,\omega')=\mathcal{M}_{21}(\omega,\omega')=\pm\frac{1}{2}\;,
\end{eqnarray}
where  $\omega$ represents the separate state $|gg\rangle$ (or $|ee\rangle$) and $\omega'$ denotes the entangled states $|\psi^{\pm}\rangle$, or vice versa.
Hence the explicit result for the contribution of vacuum fluctuations to the rate of change of the atomic energy in Eq.~(\ref{vacuum fluctuations1}) is
\begin{eqnarray}\label{total rate1}
\bigg\langle \frac{d H_A}{d\tau} \bigg\rangle_{tot}=&-&\frac{\lambda^2}{2\pi}\bigg[\sum_{\omega>\omega'}(\triangle\omega)^2
\bigg(1+\frac{1}{e^{2\pi\kappa\triangle \omega}-1}\bigg)\nonumber\\
&\times&\bigg(1\pm f(\triangle \omega,L/2)\bigg)
\nonumber\\
&-&\sum_{\omega<\omega'}(\triangle\omega)^2\frac{1}{e^{2\pi\kappa|\triangle \omega|}-1}
\bigg(1\pm f(\triangle \omega,L/2)\bigg)\bigg],\nonumber\\
\end{eqnarray}
where the (minus) plus sign refers to the (anti)symmetric Bell state ($|\psi^{-}\rangle$)  $|\psi^{+}\rangle$ and $f(\triangle \omega,L/2)=\frac{\sin[2\kappa\triangle \omega\sinh^{-1}(L/2\kappa)]}{L\triangle \omega\sqrt{1+L^2/(2\kappa)^2}}$.
It is worth noting that the rate of variation of atomic energy is modified by an extra term $f(\triangle \omega,L/2)$ as compared to that of an atom in de Sitter spacetime \cite{Yu2}. Such a function $f(\triangle \omega,L/2)$ quantifies the effect of the cross correlations mediated by the scalar field on the quantum entanglement between two atoms.  This is in sharp contrast with the case of an atom \cite{Yu2} where the effect of the spacetime curvature on the contributions of vacuum fluctuations and radiation reaction is only a ``thermal" correction with temperature $T=1/(2\pi\kappa)$. Therefore, with these differences between the two-atom system and an atom in de Sitter spacetime, we may find new effects that allows us to get the difference between the de Sitter spacetime and the thermal Minkowski spacetime.

\emph{Firstly, let us investigate the behavior of the rate of variation of energy between two atoms in de Sitter spacetime in the well known extreme cases. We identify a characteristic length scale $\kappa$ \cite{Tian} which is a function of $r$, and that the trajectories considered are not geodesics (unless $r=0$) but have curvature determined by $r$.} In the case of that the distance between two atoms are smaller than $\kappa$, it is possible to find a local inertial frame where the linear susceptibility of field is well described in Minkowski spacetime. In the other case when the interatomic distances are larger than $\kappa$, the curvature of de Sitter spacetime may have a nontrivial character.
Thus, in the limit of  $L\ll\kappa$, one has
\begin{widetext}
\begin{eqnarray}\label{smaller case}
\bigg\langle \frac{d H_A}{d\tau} \bigg\rangle_{tot}&\approx&-\frac{\lambda^2}{2\pi}\sum_{\omega>\omega'}(\triangle\omega)^2
\bigg(1+\frac{1}{e^{2\pi\kappa\triangle \omega}-1}\bigg)\bigg[1\pm \frac{1}{L\triangle \omega}
\sin(L\triangle \omega)\bigg]
\nonumber\\
&&+\frac{\lambda^2}{2\pi}\sum_{\omega<\omega'}(\triangle\omega)^2\frac{1}{e^{2\pi\kappa|\triangle \omega|}-1}
\bigg[1\pm \frac{1}{L\triangle \omega}
\sin(L\triangle \omega )\bigg],
\end{eqnarray}
while for $L\gg \kappa$, i.e., \emph{when the atoms are far apart compared to the radius of curvature of their worldlines}, we get
\begin{eqnarray}\label{larger case}
\bigg\langle \frac{d H_A}{d\tau} \bigg\rangle_{tot}&\approx&-\frac{\lambda^2}{2\pi}\sum_{\omega>\omega'}(\triangle\omega)^2
\bigg(1+\frac{1}{e^{2\pi\kappa\triangle \omega}-1}\bigg)\bigg[1\pm \frac{2\kappa}{L^2\triangle \omega}
\sin\bigg(2\kappa\triangle \omega \log\bigg(\frac{L}{\kappa}\bigg)\bigg)\bigg]
\nonumber\\
&&+\frac{\lambda^2}{2\pi}\sum_{\omega<\omega'}(\triangle\omega)^2\frac{1}{e^{2\pi\kappa|\triangle \omega|}-1}
\bigg[1\pm \frac{2\kappa}{L^2\triangle \omega}
\sin\bigg(2\kappa\triangle \omega \log\bigg(\frac{L}{\kappa}\bigg)\bigg)\bigg].
\end{eqnarray}
\end{widetext}
Let us note that when the distance between two atoms is smaller than the characteristic length scale $(L\ll\kappa)$, the function $f(\triangle \omega,L/2)$ in Eq. (\ref{total rate1}) varies with the interatomic distance as $1/L$ which is shown in Eq.~(\ref{smaller case}). However, for the case of that the interatomic distance is larger than the characteristic length scale given by Eq.~(\ref{larger case}), i.e., $L\gg\kappa$, the cross correlations behave with the interatomic distance as $1/L^2$.

It is worth to compare the above results with that in the thermal Minkowski spacetime scenario. Let us consider the rate of variation of energy between the two-atom system in Minkowski spacetime. In this case, the quantum system is composed by two identical two-level atoms interacting with a massless scalar field in a thermal bath of temperature which is described by $T=1/(2\pi\kappa$).
When the distance between the two static atoms is $ L=2r\sin(\Delta\theta/2)$ in a thermal bath of Minkowski spacetime, then the field correlation functions can be written as
\begin{eqnarray}
&&G^{+}(x_a(\tau),x_b(\tau))\nonumber\\
&&\;\;\;\;\;\;=-\frac{1}{4\pi^2}\sum^{\infty}_{n=-\infty}\frac{1}{(\Delta\tau-in/T-i\epsilon)^2},
\;\;\;\;\;\;\;\;\;(a=b),\nonumber\\
&&G^{+}(x_a(\tau),x_b(\tau'))\nonumber\\
&&\;\;\;\;\;\;=-\frac{1}{4\pi^2}\sum^{\infty}_{n=-\infty}\frac{1}{(\Delta \tau-in/T-i\epsilon)^2-L^2},\;\;(a\neq b),\nonumber\\
\end{eqnarray}
where $\Delta \tau=t-t'$ with $t$ being the proper time of the static atoms in flat spacetime and $\epsilon$ is an infinitesimal constant. According to the definition of statistical functions in Eqs.~(\ref{field1})-(\ref{atom2}), we can obtain the contributions of thermal fluctuations and the radiation reaction to the rate of change of the atomic energy, for sufficiently large $\Delta \tau$, which are given by
\begin{eqnarray}\label{thermal fluctuations3}
\bigg\langle \frac{d H_A}{d\tau} \bigg\rangle_{VF}=&-&\frac{\lambda^2}{4\pi}\bigg[\sum_{\omega>\omega'}(\triangle\omega)^2
\bigg(1+\frac{2}{e^{\triangle \omega/T}-1}\bigg)\nonumber\\
&\times&\bigg(1\pm \frac{\sin[L\triangle\omega]}{L \triangle\omega}\bigg)
\nonumber\\
&-&\sum_{\omega<\omega'}(\triangle\omega)^2\bigg(1+\frac{2}{e^{|\triangle \omega|/T}-1}\bigg)\nonumber\\
&\times&\bigg(1\pm \frac{\sin[L\triangle\omega]}{L \triangle\omega}\bigg)\bigg],
\end{eqnarray}
and
\begin{eqnarray}\label{radiation reaction2}
\bigg\langle \frac{d H_A}{d\tau} \bigg\rangle_{RR}=&-&\frac{\lambda^2}{4\pi}\bigg[\sum_{\omega>\omega'}(\triangle\omega)^2\bigg(1\pm \frac{\sin[L\triangle\omega]}{L \triangle\omega}\bigg)\nonumber\\
&+&\sum_{\omega<\omega'}(\triangle\omega)^2
\bigg(1\pm \frac{\sin[L\triangle\omega]}{L \triangle\omega}\bigg)\bigg].
\end{eqnarray}
Adding the contributions of thermal fluctuations (\ref{thermal fluctuations3}) and the radiation reaction (\ref{radiation reaction2}), one can get the total rate of change of the atomic energy
\begin{eqnarray}\label{total rate2}
\bigg\langle \frac{d H_A}{d\tau} \bigg\rangle_{tot}=
&-&\frac{\lambda^2}{2\pi}\sum_{\omega>\omega'}(\triangle\omega)^2\bigg(1+\frac{1}{e^{\triangle \omega/T}-1}\bigg)\nonumber\\
&\times&\bigg(1\pm \frac{\sin[L\triangle\omega]}{L \triangle\omega}\bigg)\nonumber\\
&+&\frac{\lambda^2}{2\pi}\sum_{\omega<\omega'}(\triangle\omega)^2
\frac{1}{e^{|\triangle \omega|/T}-1}\bigg(1\pm \frac{\sin[L\triangle\omega]}{L \triangle\omega}\bigg),\nonumber\\
\end{eqnarray}
where the (minus) plus sign refers to the (anti)symmetric Bell state ($|\psi^{-}\rangle$)  $|\psi^{+}\rangle$. Hence, we note that for the case $L\ll\kappa$ in de Sitter spacetime, the rate of variation of atomic energy given by Eq.~(\ref{smaller case}) recovers to the result obtained in Eq. (\ref{total rate2}) for a two-atom system in the thermal Minkowski spacetime.

Next, we would like to analyze the behavior of the total rate of variation of atomic energy in terms of the distance $L$ between two atoms, which is associated with the transition from the separate state to the Bell state $|\psi^\pm\rangle$ in de Sitter spacetime and thermal Minkowski spacetime respectively. This implies that we are investigating the generation of entanglement in these two different spacetime.

\begin{figure}[htbp]
\centering
\includegraphics[height=2.0in,width=3.0in]{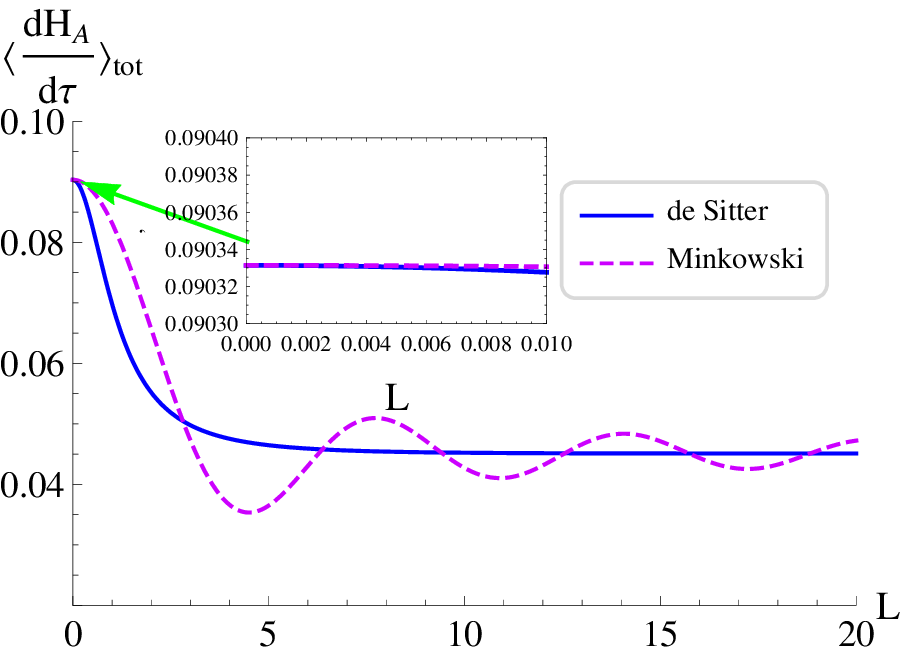}
\caption{ (color online).  The total rate of variation of atomic energy is associated with the transition  $|gg\rangle\rightarrow|\psi^+\rangle$ as functions of the parameter $L$. Solid line corresponds to the de Sitter spacetime, and dashed line to the thermal Minkowski spacetime, respectively. We choose $\gamma_0=1$, and $\kappa=0.5$. Here, $L$, $\gamma_0$ and $\kappa$ are
in the unit of $1/\omega_0$.}\label{picture1}
\end{figure}

On the one hand, we start considering the transition from the ground state to the Bell state via an absorption process, i.e., $|gg\rangle\rightarrow|\psi^\pm\rangle$ with the energy gap $\triangle\omega=-\omega_0<0$. For the case of $|gg\rangle\rightarrow|\psi^+\rangle$, according to the expressions of Eqs.~(\ref{total rate1}) and (\ref{total rate2}), in Fig. \ref{picture1} it shows the behaviors of the total rate of variation of atomic energy with respect to the interatomic distance $L$ in de Sitter spacetime and in the thermal Minkowski spacetime, for fixed $\kappa=0.5$ and $\gamma_0=1$ in the unit of $1/\omega_0$. Here, $\gamma_0=\frac{\lambda^2\omega_0}{2\pi}$ is the spontaneous emission rate. It is easy to find that $\langle d H_A/d\tau\rangle_{tot}>0$ for both spacetime, which means that the entangled state can be created from ground-state atoms via absorption process all the time.
We also can see from the plot that when the distance between the atoms is smaller than $\kappa$, i.e., for small separation between the atoms, the curve of this rate in de Sitter spacetime is almost coincided with that in the thermal Minkowski spacetime, which means that all the laws of physics in de Sitter spacetime recover to that in flat spacetime. Moreover, we are interested in noting that when the atoms are very near each other, there is an increase of the rate of variation of atomic energy induced by the influence of cross correlation, which means that the cross correlations create a constructive interference. However, as the distance between the atoms increases, the rate of variation of atomic energy for the de Sitter spacetime case and that for the thermal Minkowski spacetime case decrease differently. This is due to the fact that when the distance between two atoms is very large, i.e., the two-atom system is far apart compared to the radius of curvature of their worldlines, the constructive interference in de Sitter spacetime vanishes more quickly with power law $1/L^2$, while for the thermal Minkowski spacetime case, the constructive interference fades away in an oscillatory with power law $1/L$. \emph{More importantly, to understand the difference behaviors plotted in Fig. \ref{picture1}, it is significant to realize that the correlations in (\ref{larger case}), compared to those in (\ref{smaller case}), not only fall off faster, but also have their oscillations greatly and increasingly slowed down, because of the logarithm.}
This tells us that for short distance between the atoms the generation of entanglement via absorption process is magnified highly in comparison with the case in which the distance between the atoms is very large.

\begin{figure}[htbp]
\centering
\includegraphics[height=2.0in,width=3.0in]{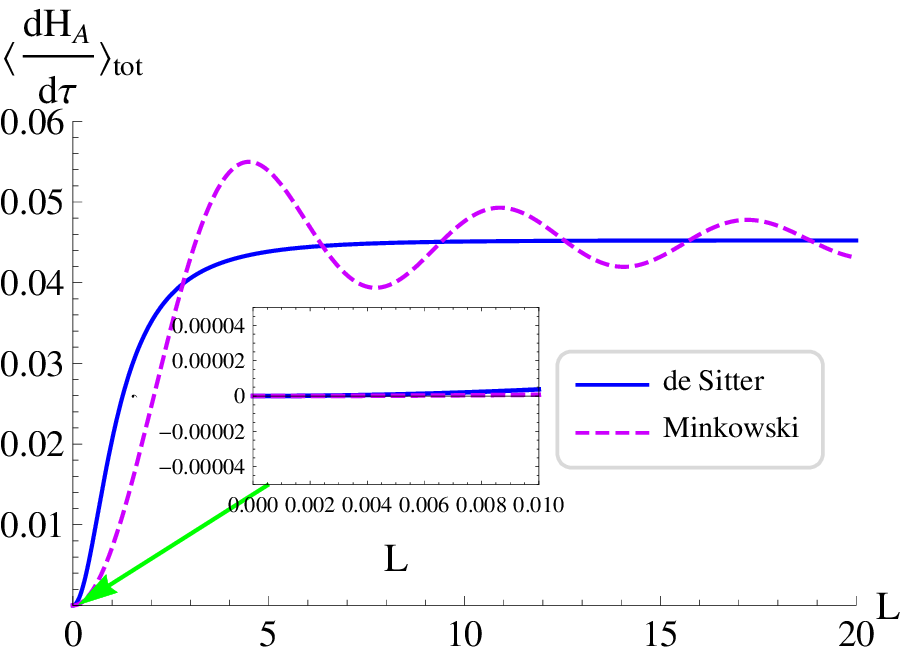}
\caption{ (color online).  The total rate of variation of atomic energy is associated with the transition $|gg\rangle\rightarrow|\psi^-\rangle$ as functions of the parameter $L$. Solid line corresponds to the de Sitter spacetime, and dashed line to the thermal Minkowski spacetime, respectively. We choose $\gamma_0=1$, and $\kappa=0.5$. Here, $L$, $\gamma_0$ and $\kappa$ are
in the unit of $1/\omega_0$.}\label{picture2}
\end{figure}

For the other case of the generation of the antisymmetric Bell state via absorption process, we have $|gg\rangle\rightarrow|\psi^-\rangle$. In Fig.~\ref{picture2} we plot the total rate of variation of energy between two atoms, in de Sitter spacetime and thermal Minkowski spacetime respectively, as functions of the interatomic distance $L$ for fixed $\kappa=0.5$ and $\gamma_0=1$. We note that for small distance between the atoms, i.e., $L\ll\kappa$, our results for the de Sitter spacetime behave the same as those of the thermal Minkowski spacetime. Besides, it is interesting to find that as the atoms approach each other, one has $\langle H_A/d\tau\rangle_{tot}\rightarrow0$, which means that the entangled state $|\psi^-\rangle$ cannot be created from the ground state $|gg\rangle$ as a consequence of the destructive interference of quantum correlations between the atoms. In this respect, we can recover the well known result as considered in Ref. \cite{Ficek2}, which states that for atoms confined in a region much smaller than the optical wavelength, one regards the antisymmetric Bell state as a decoherence-free state.
However, when the distance between the atoms is very large, i.e., $L\gg\kappa$, the rate of variation of energy increases differently, because the interference in de Sitter spacetime will vanish more quickly with power law $1/L^2$ than that for the thermal Minkowski spacetime case with power law $1/L$.
Therefore, the generation of quantum entanglement via absorption process is largely inhibited when the distance between the atoms is small compared with that for large distance.

\begin{figure}[htbp]
\centering
\includegraphics[height=2.0in,width=3.0in]{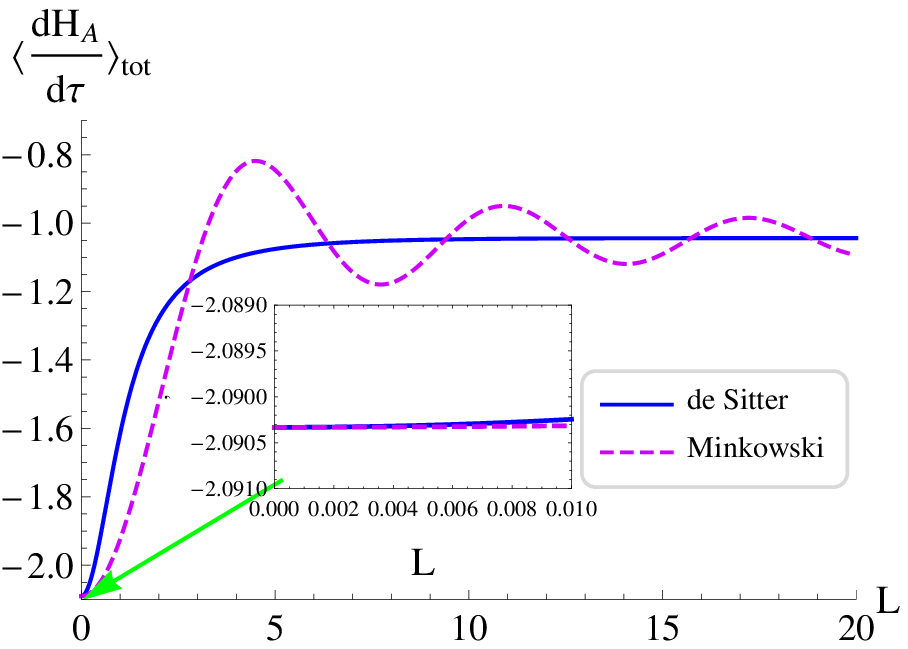}
\caption{ (color online).  The total rate of variation of atomic energy related to the transition $|ee\rangle\rightarrow|\psi^+\rangle$, is the functions of the parameter $L$. Solid line corresponds to the de Sitter spacetime, and dashed line to the thermal Minkowski spacetime, respectively. We choose $\gamma_0=1$, and $\kappa=0.5$. Here, $L$, $\gamma_0$ and $\kappa$ are
in the unit of $1/\omega_0$.}\label{picture3}
\end{figure}

On the other hand, we would like to discuss the transition $|ee\rangle\rightarrow|\psi^\pm\rangle$ in de Sitter spacetime and thermal Minkowski spacetime, and study the generation of entanglement via emission process. With this respect, we get $\triangle\omega=\omega_0>0$. Firstly,
we consider the case of the generation of the symmetric Bell state via spontaneous emission process, i.e., $|ee\rangle\rightarrow|\psi^+\rangle$. Then we plot the rate of variation of atomic energy as a function of $L$ in Fig.~\ref{picture3}, for fixed $\kappa=0.5$ and $\gamma_0=1$ in de Sitter spacetime and the thermal Minkowski spacetime respectively. We can see from the Fig.~\ref{picture3} that $\langle H_A/d\tau\rangle_{tot}<0$ is independent of how far the atoms are separated for both spacetime, which implies that the entangled state always can be generated via emission process. Again, we find that as the distance between the atoms is small $(L\ll\kappa)$, the results obtained in de Sitter spacetime are the same with that in the thermal Minkowski spacetime. In addition, as the atoms very close to each other, both effects of fluctuations and the radiation reaction lead to a violently loss of atomic energy via emission process, because of the constructive interference of cross correlations between the atoms. It is also worth noting that for a large separation between the atoms $(L\gg\kappa)$, the loss of atomic energy via emission process is weakened with different behaviors in de Sitter spacetime and thermal Minkowski spacetime. The reason is that the constructive interference in de Sitter case decays quickly as a consequence of the effect of the curvature of de Sitter spacetime, while this interference dies off in an oscillatory manner in flat spacetime. It shows that there is a greatly enhanced generation of entanglement between the atoms via emission process as the atoms approach each other.

\begin{figure}[htbp]
\centering
\includegraphics[height=2.0in,width=3.0in]{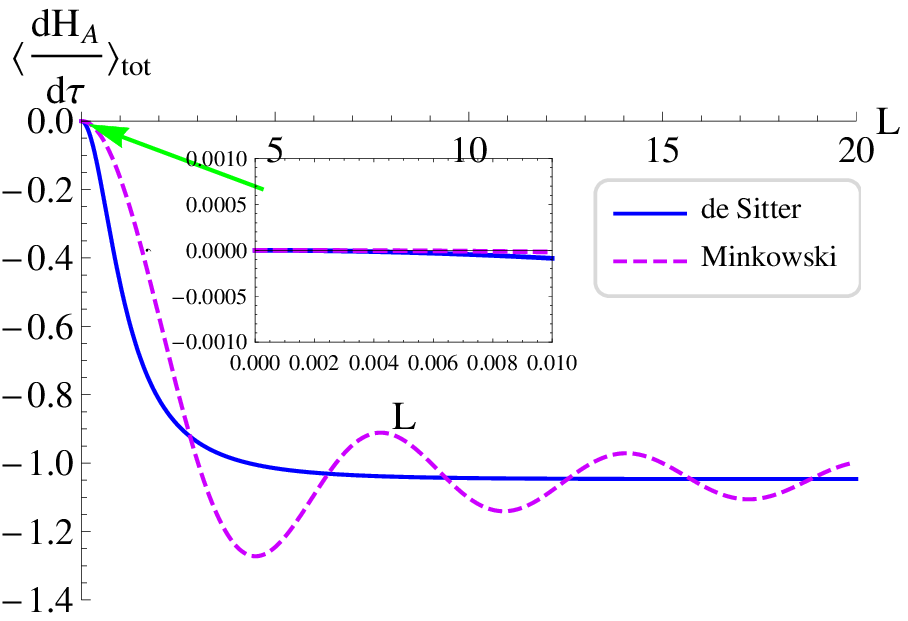}
\caption{ (color online).  The total rate of variation of atomic energy related to the transition $|ee\rangle\rightarrow|\psi^-\rangle$, is the functions of the parameter $L$. Solid line corresponds to the de Sitter spacetime, and dashed line to the thermal Minkowski spacetime, respectively. We choose $\gamma_0=1$, and $\kappa=0.5$. Here, $L$, $\gamma_0$ and $\kappa$ are
in the unit of $1/\omega_0$.}\label{picture4}
\end{figure}

For the case $|ee\rangle\rightarrow|\psi^-\rangle$, we illustrate the behavior of the rate of variation of atomic energy as functions of the interatomic distance $L$ for fixed $\kappa=0.5$ and $\gamma_0=1$ in Fig.~\ref{picture4}. We note that the de Sitter spacetime and the thermal Minkowski spacetime share the same properties when the distance between the atoms is small $(L\ll\kappa)$. Moreover, we remark that for the atoms are very near to each other, one gets $\langle d H_A/d\tau\rangle_{tot}\rightarrow0$ for the antisymmetric state, which represents that the Bell state $|\psi^-\rangle$ is stable via emission processes due to destructive interference of cross correlations between the atoms. Besides, as the interatomic distance increases, the absolute values of the rate of variation of atomic energy increase differently for the case of de Sitter spacetime and the thermal Minkowski spacetime, which implies that the loss of atomic energy via emission process grows differently. This is resulted from that the destructive interference in de Sitter spacetime disappears quickly, but for the thermal Minkowski spacetime case it dies off in an oscillatory manner. Therefore, it tells us that the generation of entanglement via emission process is largely suppressed when the atoms are near each other.
Incidentally, when we study the properties of the rate of variation of atomic energy associated with the transition $|\psi^\pm\rangle\rightarrow|gg\rangle$ and $|\psi^\pm\rangle\rightarrow|ee\rangle$, i.e., degradation of quantum entanglement,  the results obtained in de Sitter spacetime and thermal Minkowski spacetime are qualitatively similar to those discussed above as expected.

\section{conclusion}
We have studied the radiative processes of two-level atoms interacting with a conformally coupled massless scalar field prepared in the de Sitter-invariant vacuum. Employing the formalism developed by Dalibard, Dupont-Roc and Cohen-Tannoudji, we investigated the distinct contributions of vacuum fluctuations and the radiation reaction to the quantum entanglement between two identical static atoms in de Sitter spacetime.
We have shown that when the distance between the atoms smaller than the characteristic length scale, the rate of variation of atomic energy in de Sitter spacetime is similar to that in thermal Minkowski spacetime. However, when beyond the characteristic length scale, this rate in both spacetime behaves differently. In this paper, two different entangled states, the symmetric Bell state and the antisymmetric Bell state, have been studied.

For the symmetric Bell state, it can always be generated even if they are initially prepared in a separable state. Moreover, we found that the generation or degradation entanglement are highly magnified when the atoms are near to each other, as a consequence of that the cross correlation between the atoms generates a constructive interference for short distance and then this interference decreases for large separation between the atoms. However, when the distance between the atoms is very large, the interference in de Sitter spacetime vanishes quickly, while for the thermal Minkowski spacetime, it dies off in an oscillatory manner.

On the other hand, for the antisymmetric Bell state case, the generation or degradation entanglement will be largely suppressed when the atoms approach each other in comparison with the case in which the interatomic distance is very large. That is, when the atoms are very near to each other, the antisymmetric Bell state is stable with respect to radiation process. This is because for small distance, the quantum correlation between the atoms creates destructive interference which disappears for large distance. In addition, for a very large separation between the atoms, this interference in de Sitter spacetime vanishes more quickly than that in thermal Minkowski spacetime.

\begin{acknowledgments}
This work is supported by the  National Natural Science Foundation
of China under Grant Nos. 11475061 and 11675052.
X. Liu thanks for the Hunan Provincial Innovation Foundation For Postgraduate under Grant No. CX2017B175.
Z. Tian was supported by the BK21 Plus Program (21A20131111123) funded by the Ministry of Education (MOE, Korea) and National Research Foundation of Korea (NRF).
\end{acknowledgments}

\end{document}